\documentclass[12pt]{iopart}
\usepackage{graphicx}
\begin{document}
\jl{1}
\title[The probability distribution function of the local energy in Ising spin
glass]{The nature of the probability distribution function of
the local energy in Ising spin glass}

\author{Hidetsugu 
Kitatani\footnote[3]{E-mail address:kitatani@vos.nagaokaut.ac.jp},
and Akira Aoki}

\address{Department of Electrical Engineering, Nagaoka University of
Technology, Nagaoka, Niigata 940-2188, Japan}

\begin{abstract}
The nature of the probability distribution function 
of the local energy in the $\pm J$ Ising model
has been investigated.
At finite temperature, it has been derived that the probability
distribution function
must satisfy several  relations
at $p=1/2$ ($p$ is the concentration
of the ferromagnetic bond) and at Nishimori-line, respectively
on any lattice in any dimension.
They relate the probability distribution function
corresponding to the local energy lower than $-\tanh (K)$
with that corresponding to the local enegy
greater than $-\tanh (K)$. ($K$ is the 
inverse temperature.)
The present results at Nishimori-line are, in a sense,
generalization of
Nishimori's result about the internal energy obtained by 
 the local gauge transformation.
Moreover, from the numerical calculation
in the two-dimentional $\pm J$ Ising model,
it is found that, in a certain
temperature region,
the probability distribution function of the local energy
has several peaks which are related to
 the  patterns of frustration
around a  bond of the lattice.
\end{abstract}

%
%
\pacs{75.50.Lk,64.60.Cn,05.50.+q}

\section{Introduction}
To elucidate the nature  of random spin systems,
especially spin glass systems,
has been a subject of a long-standing interest[1-21].
In the random spin systems, when we take one sample,
namely, one bond configuration, 
the local energies of interacting bonds take various values,
which change as the sample
changes. There have been many works about the energy
of spin glass systems, for example,  about the
ground state energy[7-9], energy barrier[10-12], energy landscape
[13-15],
low-energy excitation[16,17] for various spin glass systems.

On the other hand, 
Nishimori[18]
 derived several rigorous results  at
a special line in the phase diagram of spin glass systems, which
has now been called "Nishimori-line".
The results were derived mainly by the use of
the local gauge transformation, so that
they hold for various spin glass models
on any lattice in any dimension.
The exact internal energy and
the upper bound of the specific heat
at Nishimori-line were derived.
Moreover, a nature of the correlation function
was derived, from
which it follows that
the ferromagnetic order parameter and the
spin glass order parameter coincide with 
each other at Nisimori-line, and the ferromagnetic phase boundary
has some restriction in the phase diagram[18-20].
Furthermore, it was derived  that the correlation-function
distribution function must satisfy a certain relation
at Nishimori-line[21].

In this paper,
we investigate the nature of the probability
distribution function of the local
energy in the $\pm J$ Ising model.
It has been derived that, at finite temperature,
 the probability distribution
function mentioned above must satisfy several
relations at $p=1/2$ and at Nishimori-line, respectively, 
on any lattice in any dimension.
They relate the probability distribution function
corresponding to the local energy lower than $-\tanh (K)$
with that corresponding to the local enegy
greater than $-\tanh (K)$.
The present results  at Nishimori-line  are, in a sense,
generalization of Nishimori's result about
the internal energy[18], since
the probability distribution function has more
information than only the average value.

Moreover, we have numerically calculated
the probability distribution functions
of the local energy at $p=1/2$ and at Nishimori-line 
in the two-dimensional $\pm J$ Ising model, from which
we have found that,
in a certain temperature region, they have several peaks
which are related to  the
patterns of frustration around a bond
of the lattice.

\section{The probability distribution function of the local energy}
We consider the  $\pm J$ Ising model,
where the dimension of the lattice, the lattice
structure and the range of the interactions may be arbitrary.
The Hamiltonian  is written as follows:
\begin{equation}
 {\cal H} = -\sum_{(ij)}\tau_{ij}S_{i}S_{j}, 
\end{equation}
where $S_{i}=\pm 1$, and the
summation of $(ij)$ runs over all the interacting pairs.
 Each $\tau_{ij}$ is determined according to the following
probability distribution:
\begin{equation}
    P(\tau_{ij}) = p\delta (\tau_{ij}-1)+(1-p)\delta (\tau_{ij}+1).
\end{equation}
In this paper, we put that $J=1$ and $k_{\rm B}=1$ ($k_{\rm B}$ is the Boltzmann constant).

Now, we denote the local energy of the interacting bond $(ij)$ in
a given bond configuration, $\{\tau \}$, as $e_{ij}(K)$:
\begin{equation}
   e_{ij}(K)=-<\tau_{ij}S_{i}S_{j}>_{K},
\end{equation}
where $<\cdots >_{K}$ denotes the thermal average
at temperature, $T=1/K$.
Then, the probability distribution function
of the local energy, $P_{e}(x,K,K_{p})$, can be written as
\begin{equation}
P_{e}(x,K,K_{p})=[\delta (x-e_{ij}(K))]_{K_{p}},
\end{equation}
where $[\cdots ]_{K_{p}}$ denotes the configurational average at the
ferromagnetic bond concentration, $p$.
(We define $K_{p}$ as $\exp (2K_{p})=p/(1-p)$.)
In this paper, we do not investigate the nature
of $P_{e}(x,K,K_{p})$ directly,
but
investigate that of a slightly different form,
namely,
\begin{equation}
P_{e}^{(2)}(x,K,K_{p})=[\delta (x-(\cosh (2K)+\sinh (2K)e_{ij}(K)))]_{K_{p}},
\end{equation}
which is related to $P_{e}(x,K,K_{p})$ as
\begin{equation}
P_{e}(x,K,K_{p})=\sinh (2K)P_{e}^{(2)}(\cosh (2K)+\sinh (2K)x,K,K_{p}).
\end{equation}
Here, it is noted that $P_{e}(x,K,K_{p})$ may take non-zero value
in the region, $-1\leq x \leq 1$, so that $P_{e}^{(2)}(x,K,K_{p})$
may take non-zero value in the region, $\exp (-2K) \leq x \leq \exp (2K)$.

\section{The nature of the probability distribution function of the local energy}
In this section, we investigate the nature of the probabilty
distribution function of the local energy, $P_{e}^{(2)}(x,K,K_{p})$, at $p=1/2$
and at Nishimori-line, respectively.

Firstly, we can derive the following identity:
\begin{eqnarray}
\lefteqn{
P_{e}^{(2)}(x,K,K_{p})
} \nonumber \\
{
=[(\cosh(2K_{p})+\sinh(2K_{p})e_{ij}(K_{p}))
\delta (x-\frac {1}{\cosh(2K)+\sinh(2K)e_{ij}(K)})]_{K_{p}}.
} \nonumber \\
\end{eqnarray}
(For the detailed derivation of equation (7), see the appendix.)

Next, we integrate both terms of equation (7)  from
$a$ to $b$ ($a$ and $b$ are arbitrary real numbers, which satisfy 
$\exp (-2K)\leq a<b\leq \exp (2K)$.):
namely,
\begin{eqnarray}
\lefteqn{
\int_{a}^{b}P_{e}^{(2)}(x,K,K_{p})dx
} \nonumber \\
\lefteqn{
=\int_{a}^{b}[(\cosh(2K_{p})+\sinh(2K_{p})e_{ij}(K_{p}))
\delta (x-\frac {1}{\cosh(2K)+\sinh(2K)e_{ij}(K)}]_{K_{p}}dx
} \nonumber \\
\lefteqn{
=\int_{1/b}^{1/a}
[(\cosh(2K_{p})+\sinh(2K_{p})e_{ij}(K_{p}))
\delta (\frac {1}{x}-\frac {1}{\cosh(2K)+\sinh(2K)e_{ij}(K)}]_{K_{p}}
\frac {dx}{x^{2}}
} \nonumber \\
{
=\int_{1/b}^{1/a}
[(\cosh(2K_{p})+\sinh(2K_{p})e_{ij}(K_{p}))
\delta (x-(\cosh(2K)+\sinh(2K)e_{ij}(K))]_{K_{p}}dx,
} \nonumber \\
\end{eqnarray}
where we use the following property of $\delta$-function
to derive the last term:
\begin{equation}
\delta(\frac {1}{x}-\frac {1}{a})=x^{2}\delta(x-a).
\end{equation}
Equation (8) is the basic identity, which we use
to derive the properties of the probability distribution function
of the local energy.

First, we investigate the case, $p=1/2$.
At $p=1/2$,  
 $K_{p}=0$, so that equation (8) becomes
\begin{eqnarray}
\lefteqn{
\int_{a}^{b}P_{e}^{(2)}(x,K,0)dx
} \nonumber \\
\lefteqn{
=\int_{1/b}^{1/a}
[\delta (x-(\cosh(2K)+\sinh(2K)e_{ij}(K)))]_{0}dx
} \nonumber \\
=\int_{1/b}^{1/a}
P_{e}^{(2)}(x,K,0)dx
\end{eqnarray}
Changing the variable $x$ into $x^{-1}$ of rhs of equation (10), we also
obtain
\begin{equation}
\int_{a}^{b}P_{e}^{(2)}(x,K,0)dx=\int_{a}^{b}x^{-2}P_{e}^{(2)}(x^{-1},K,0)dx.
\end{equation}
For investigating the direct form of $P_{e}^{(2)}(x,K,K_{p})$,
we define the following averaged distribution function, $\{P_{e}^{(2)}(x,K,K_{p}\}
_{\Delta}$:
\begin{equation}
\{P_{e}^{(2)}(x,K,K_{p}) \}_{\Delta}=\frac {1}{\Delta}\int_{x}^{x+\Delta}P_{e}^{(2)}(x,K,K_{p})dx,
\end{equation}
where $\Delta$ is an arbitrary  finite value which satisfies $x+\Delta \leq
 \exp (2K)$.
Then, equation (11) can be written as:
\begin{equation}
\{P_{e}^{(2)}(x,K,0)\}_{\Delta}=\{x^{-2}P_{e}^{(2)}(x^{-1},K,0)\}_{\Delta}.
\end{equation}
We can take the value, $\Delta$, as an arbitrarily small but
finite value.
In this sense, we can conclude that the 
value of $P_{e}^{(2)}(x,K,0)$ coincides with that of $x^{-2}P_{e}^{(2)}(x^{-1},K,0)$
at any finite temperature.
Equations (10), (11) and (13) are the main results at $p=1/2$.
In a word, these equations are the conditions which
relates the value of the
probability distribution function at $x$ with that at $1/x$. 
As the temperature changes, the distribution function
of the local energy itself may change,
however,
the distribution function  must satisfy the above
relations 
at any finite temperature on any lattice in any dimension.
Using $P_{e}(x,K,K_{p})$, equation (13) is written by  the following form:
\begin{eqnarray}
\lefteqn{
  \{ P_{e}(x-(-\tanh (K)),K,0)\}_{\Delta}
}  \nonumber  \\
{
=\{ \frac {1}{(1+\sinh(2K)(x-(-\tanh (K))))^{2}}
P_{e}(-\frac {x-(-\tanh (K))}{1+\sinh (2k)(x-(-\tanh (K)))},K,0)\}_{\Delta},
}  \nonumber \\
\end{eqnarray}
which becomes a rather complicated form, from which, however,
we can see that equation (14)
is a condition which relates the probability distribution function
corresponding to the local energy lower than $-\tanh (K)$
with that corresponding to the local energy
greater than $-\tanh (K)$.

Next, we investigate the nature of the probability
 distribution function
of the local energy at Nishimori-line.
At Nishimori-line, $K_{p}=K$,  so that equation (8)  becomes
\begin{eqnarray}
\lefteqn{
\int_{a}^{b}P_{e}^{(2)}(x,K,K)dx
} \nonumber \\
\lefteqn{
=\int_{1/b}^{1/a}
[(\cosh(2K)+\sinh(2K)e_{ij}(K))
\delta (x-(\cosh(2K)+\sinh(2K)e_{ij}(K))]_{K}dx
} \nonumber \\
\lefteqn{
=\int_{1/b}^{1/a}
[x\delta (x-(\cosh(2K)+\sinh(2K)e_{ij}(K))]_{K}dx
} \nonumber \\
=\int_{1/b}^{1/a}xP_{e}^{(2)}(x,K,K)dx.
\end{eqnarray}
Also, we obtain
\begin{equation}
\int_{a}^{b}P_{e}^{(2)}(x,K,K)dx=\int_{a}^{b}x^{-3}P_{e}^{(2)}(x^{-1},K,K)dx.
\end{equation}
For the averaged distribution function, it can be derived that
\begin{equation}
\{P_{e}^{(2)}(x,K,K)\}_{\Delta}=\{x^{-3}P_{e}^{(2)}(x,K,K)\}_{\Delta}.
\end{equation}

\section{The property of the local energy at $p=1/2$ and at Nishimori-line}
In this section, we derive several properties of the local energy
at $p=1/2$ and at Nishimori-line, using the results of the 
preceding section.

For the configurational average of the local energy
at Nishimori-line,
from equation (15) , we obtain
\begin{eqnarray}
\lefteqn{
[\cosh (2K)+\sinh (2K)e_{ij}(K)]_{K}
} \nonumber \\
\lefteqn{
=\int_{\exp(-2K)}^{\exp(2K)}xP_{e}^{(2)}(x,K,K)dx
} \nonumber \\
=\int_{\exp(-2K)}^{\exp(2K)}P_{e}^{(2)}(x,K,K)dx=1,
\end{eqnarray}
from which, it is easily calculated that
\begin{equation}
[e_{ij}(K)]_{K}=-\tanh (K),
\end{equation}
which was first derived by Nishimori[18].
At $p=1/2$, by the similar procedure, it can be derived that
\begin{equation}
[e_{ij}(K)]_{0} \geq -\tanh (K).
\end{equation}

Next, putting $a=\exp (-2K)$ and $b=1$ in equation (10) at $p=1/2$,
we obtain
\begin{equation}
\int_{\exp(-2K)}^{1}P_{e}^{(2)}(x,K,0)dx=\int_{1}^{\exp(2K)}P_{e}^{(2)}(x,K,0)dx.
\end{equation}
Using $P_{e}(x,K,K_{p})$, equation (21) can be written as
\begin{equation}
\int_{-1}^{-\tanh(K)}P_{e}(x,K,0)dx=\int_{-\tanh(K)}^{1}P_{e}(x,K,0)dx.
\end{equation}
On the other hand, at Nishimori-line, from equation (15), we
obtain
\begin{equation}
\int_{-1}^{-\tanh(K)}P_{e}(x,K,K)dx \geq \int_{-\tanh(K)}^{1}P_{e}(x,K,K)dx.
\end{equation}

The above results are intersting, since, at Nishimori-line,
the configurational average of the local energy coincides with
$-\tanh (K)$ at any temperature, while, at $p=1/2$,
the probability that the local energy, $e_{ij}(K)$,
 takes the value smaller than
$-\tanh (K)$ always coincides with one that
$e_{ij}(K)$ takes the value larger than
$-\tanh (K)$
at any temperature.
Furthermore, these results hold on  any lattice structure
in any dimension.

\section{Numerical calculation of the probability distribution function}
Now, we show the examples how the above feature
holds in the two-dimensional square-lattice $\pm J$
Ising model with only nearest neighbour interactions.
By the transfer matrix method, we have calculated
the values of $e_{ij}(K)$ 
in $L \times L$ lattice ($L=11$) for $10^{7}$
bond configurations, from  which
we have estimated $\{P_{e}^{(2)}(x,K,K_{p})\}_{\Delta}$ at $p=1/2$ and
at
Nishimori-line for $\Delta=(\exp (2K)-\exp (-2K))/1000$.
Figures 1 and 2 are the results of the above procedure
at $K=0.75$ at $p=1/2$ and at Nisimori-line, respectively.
We have confirmed that equations (13) and (17) definitely hold
within the statistical errors.

In the figures, we can see that 
the averaged probability distribution
function, $\{P_{e}^{(2)}(x,K,K_{p})\}_{\Delta}$, has three peaks
at $p=1/2$, and one peak and two shoulders at Nishimori-line.
In the two-dimensional
square lattice $\pm J$ Ising model, the number of
the patterns of furstration of two plaquettes around
a certain bond of the lattice is three if we take 
into account only the number of frustration plaquettes, and each
bond configuration belongs to one of the three patterns.
We have confirmed that
each peak or shoulder corresponds to 
the contribution from one of the three patterns of frustration.
As the temperature decreases, each peak broadens,
and at sufficient low temperature, it is found
that $\{P_{e}^{(2)}(x,K,K_{p})\}_{\Delta}$ has only one peak.
More detailed numerical properties of the
probability distribution 
function of the local energy
 is reported
in a separated paper in the future.

\section{Conclusions}
We have investigated the nature of the probability
distribution function of the local energy
in Ising spin glass.

We have derived that, in the $\pm J$
Ising model, the probability
distribution function, $P_{e}^{(2)}(x,K,K_{p})(P_{e}(x,K,K_{p}))$,
must satisfy several relations 
at  finite temperature on any lattice in any dimension
at $p=1/2$ and
at Nishimori-line, respectively.
They relate the probability distribution function
corresponding to the local energy lower than $-\tanh (K)$
with that corresponding to the local enegy
greater than $-\tanh (K)$.
The present results at Nishimori-line are 
generalization of Nishimori's result[18] about
the internal energy, since the
probability distribution function
has more information than only the average value.

Moreover, from the numerical calculation,
we have found that, in a certain temperature region,
 the probability distribution
function of the local enargy has several peaks
which are related to the 
patterns of frustration around a certain bond
in the lattice. More
detailed numerical properties are
reported in the near future.

\ack
The authors would thank Dr. S. Kobayashi and Dr. S. Hara for fruiful
discussions.
The calculations were made on HITAC SR8000 and SR11000 at the Institue for
Solid State Physics in University of Tokyo and at University of Tokyo.

\appendix
\section{Derivation of equation (7)}
In this appendix, we briefly explain the
derivation of equation (7).

First, we denote the partition function of the system
in a given bond configuration, $\{\tau \}$ as;
\begin{equation}
Z(K,K^{'})=\sum_{\{S\}}\exp(\sum_{lm\ne ij}K\tau_{lm}S_{l}S_{m}
+K^{'}\tau_{ij}S_{i}S_{j}),
\end{equation}
where we denote the temperature at bond $(ij)$ separately.
Also, we introduce the notation,$[\cdots]_{K_{p},K_{p}^{'}}$:
\begin{eqnarray}
 \lefteqn{  [\cdots]_{K_{p},K_{p}^{'}}
} \nonumber \\
=\frac {1}{(2\cosh (K_{p}))^{N_{\rm B}-1}2\cosh (K_{p}^{'})}\sum_{\{\tau\}
}\exp({K_{p}
\sum_{lm\ne ij}\tau_{lm}+K_{p}^{'}\tau_{ij}})\cdots,
\end{eqnarray}
where $N_{\rm B}$ is the numbers of bonds, and we denote the ferromagnetic bond
concentration
of bond $(ij)$  separately.
Of course, if $K_{p}^{'}=K_{p}$, the notation mentioned above coincides with the standard one:
\begin{equation}
[\cdots]_{K_{p},K_{p}}
=[\cdots]_{K_{p}}.
\end{equation}

It is easily obtained that
\begin{equation}
\cosh (2K)+\sinh(2K)e_{ij}(K)
=\frac {Z(K,-K)}{Z(K,K)}.
\end{equation}
Then, it yields that
\begin{eqnarray}
\lefteqn{
P_{e}^{(2)}(x,K,K_{P})=[\delta(x-(\cosh(2K)+\sinh(2K)e_{ij}(K)))]_{K_{p}}
} \nonumber \\
=[\delta(x-\frac {Z(K,-K)}{Z(K,K)})]_{K_{p},K_{p}}
\end{eqnarray}
The last term of the above equation is invariant when we 
change the sign of $K$ and  $K_{p}$ at 
bond $(ij)$ simultaneously, since it means that we just take
the summation of $\tau_{ij}(=\pm 1)$ reversely; namely,
\begin{eqnarray}
\lefteqn{
P_{e}^{(2)}(x,K,K_{p})
} \nonumber \\
\lefteqn{
=[\delta(x-\frac {Z(K,K)}{Z(K,-K)})]_{K_{p},-K_{p}}
} \nonumber \\
\lefteqn{
=[\exp(-2K_{p}\tau_{ij})\delta(x-\frac {1}{\frac {Z(K,-K)}{Z(K,K)}})]_{K_{p},K_{p}}
} \nonumber \\
\lefteqn{
=[(\cosh(2K_{p})-\sinh(2K_{p})\tau_{ij})
\delta(x-\frac {1}{\cosh(2K)+\sinh(2K)e_{ij}(K)})]_{K_{p}}
} \nonumber \\
{
=[(\cosh(2K_{p})+\sinh(2K_{p})e_{ij}(K))
\delta(x-\frac {1}{\cosh(2K)+\sinh(2K)e_{ij}(K)})]_{K_{p}},
} \nonumber \\
\end{eqnarray}
where we use the local gauge transformation to derive the last term. 
Thus,  we  obtain equation (7).

\Figures

\begin{figure}
\centerline{\includegraphics[width=12cm,height=12cm]
                    {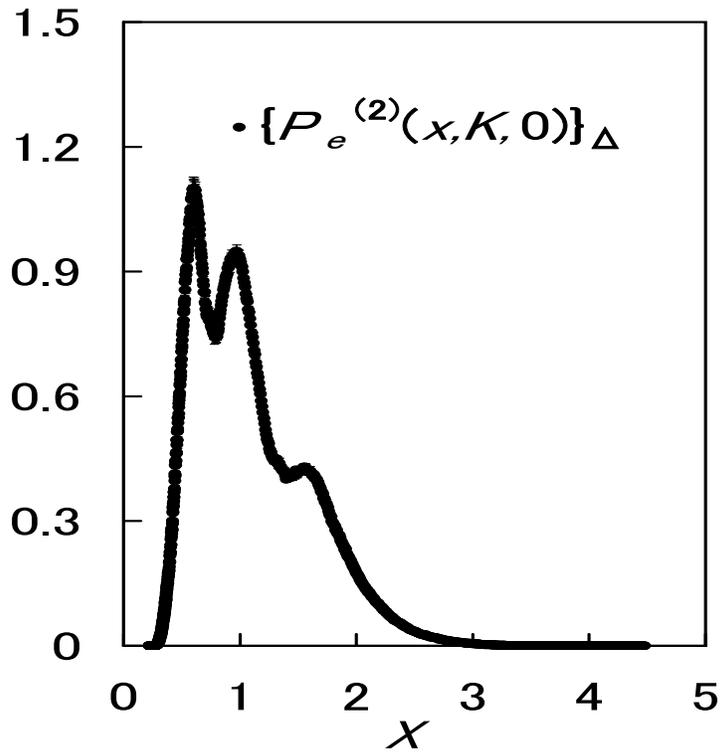}}
\caption{The averaged probability distribution
function, $\{P_{e}^{(2)}(x,K,0)\}_{\Delta}$, at $K=0.75$ at $p=1/2$.}

\end{figure}

\begin{figure}
\centerline{\includegraphics[width=12cm,height=12cm]
                    {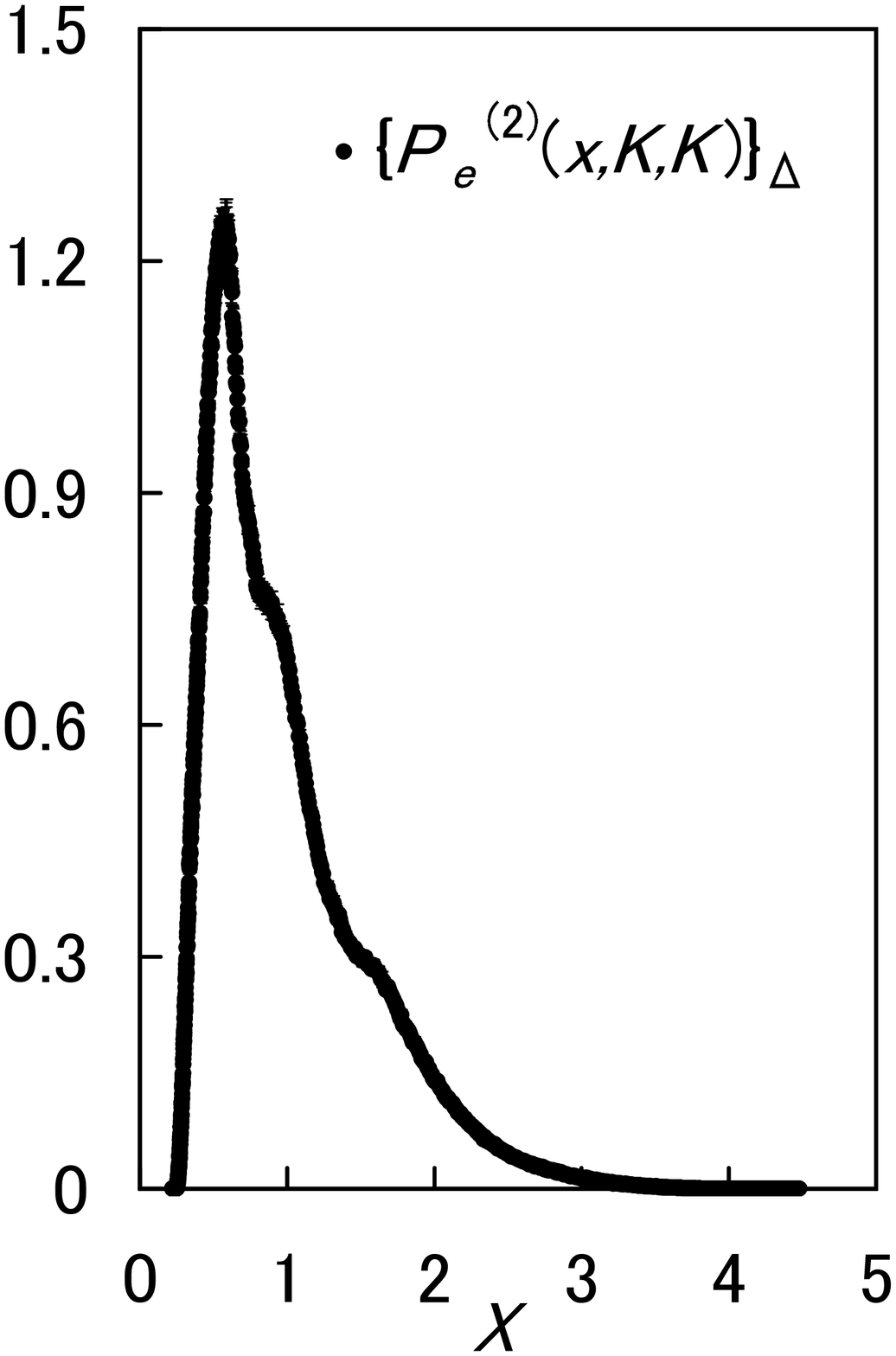}}
\caption{The averaged probability distribution function, $\{P_{e}^{(2)}(x,K,K)\}_{\Delta}$,
at $K=0.75$ at Nishimori-line.}
\end{figure}


\section*{References}
\begin{thebibliography}{100}
\bibitem{Edwards} Edwrads S F and Anderson P W 1975
{\it J. Phys. F} {\bf 5} 965
\bibitem{Sherrington} Sherrington D and Kirkpatrick S 1975
{\it Phys. Rev. Lett.} {\bf 35} 1792
\bibitem{Parisi} Parisi G 1980
{\it J. Phys. A} {\bf 13} L115
\bibitem{Bhatt} Bhatt R N and Young A P 1985
{\it J. Phys. Rev. Lett.} {\bf 54} 924
\bibitem{Ogielski} Ogielski A T and Morgenstern I 1985
{\it J. Phys. Rev. Lett.} {\bf 54} 928
\bibitem{Ballesteros} Ballesteros H G, Cruz A, Fernandez L A,
Martin-Mayor V, Pech J, Ruiz-Lorenzo J J, Tarancon A,
Tellez P Ullod C L  and Ungil C 2000
{\it Phys. Rev. B} {\bf 62} 14237
\bibitem{McMillan} Dheung H F and McMillan W L
{\it J. Phys. C} {\bf 16} (1983) 7027
\bibitem{Hartmann} Hartmann A K
{\it Physica A} {\bf 224} (1996) 480
\bibitem{Palmer} Palmer R G and Adler J
{\it Int. J. Mod. Phys. C} {\bf 10} (1999) 667
\bibitem{Rogers} Rogers G J and Moore M A
{\it J. Phys. A} {\bf 22} (1989) 1085
\bibitem{Fontanari} Fontanari J F and Stadler P F
{\it J. Phys. A} {\bf 35} (2002) 1509
\bibitem{Drossel} Drossel B and Moore M A
{\it Phys. Rev. B} {\bf 70} (2004) 064412
\bibitem{Garstecki} Garstecki P, Hoang T X and Cieplak M
{\it Phys. Rev. B} {\bf 60} (1999) 3219
\bibitem{Glotzer} Glotzer S C, Jan N and Poole P H
{\it J. Phys. Cod. Matt. } {\bf 12} (2000) 6675
\bibitem{Krawczyk} Krawczyk J and Kobe  S
{\it Physica A}{\bf 315} (2002) 1509
\bibitem{Katzgraber} Katzgraber H G, Palassini M and Young A P
{\it Phys. Rev. B} {\bf 63} (2001) 184422
\bibitem{Lamarcq} Lamarcq J, Bouchaud J P, Martin O and Mezard M
{\it Europhys. Lett. } {\bf 58} (2002) 321
\bibitem{Nishimori} Nishimori H 1981
{\it Prog. Theor. Phys.} {\bf 66} 1169
\bibitem{Kitatani} Kitatani H 1992
{\it J. Phys. Soc. Jpn.} {\bf 61} 4049
\bibitem{Ozeki} Ozeki Y and Nishimori H 1993
{\it J. Phys. A} {\bf 26} 3399
\bibitem{Queiroz} Queiroz S L A, Stinchcombe
{\it Phys. Rev. B} {\bf 68} (2003) 144414

\end{thebibliography}
\end{document}